\documentstyle[12pt,aps]{revtex}
\begin{document}
\preprint{DOE/ER/40561-295-INT96-00-152} 
\title{On the calculation of hadron form factors from\\
Euclidean Dyson-Schwinger equations}
\vspace{0.5in}
\author{M. Burkardt$^\dagger $, M.R. Frank$^\ast $, and K.L. 
Mitchell$^\ddagger$}
\address{$\dagger $ Department of Physics, New Mexico State University,
Las Cruces, NM 88003-0001}
\address{$\ast $Institute for Nuclear Theory, University of Washington, Seattle, 
WA 98195}
\address{$\ddagger $TRIUMF, 4004 Wesbrook Mall, Vancouver, British Columbia, 
Canada, V6T 2A3}
\maketitle
\vspace{0.5in}
\begin{abstract}
We apply Euclidean time methods to phenomenological
Dyson-Schwinger models
of hadrons. By performing a Fourier transform of the momentum
space correlation function to Euclidean time and by taking the
large Euclidean time limit, we project onto the lightest on-mass-shell
hadron for given quantum numbers.
The procedure, which actually resembles lattice gauge theory methods,
allows the extraction of moments of structure functions,
moments of light-cone wave functions and form factors without
{\it ad hoc} extrapolations to the on-mass-shell points.
We demonstrate the practicality of the procedure with the example of
the pion form factor.
\end{abstract}
\section{introduction}
The planned experimental program at the Thomas Jefferson National Accelerator 
Facility (TJNAF) will subject the electromagnetic structure 
of hadrons and nuclei to detailed scrutiny.  The calculation of hadron 
form factors is therefore of fundamental importance to the subsequent 
interpretation of the obtained experimental results.  However the 
nonperturbative description 
of hadron structure in the Minkowski metric, which has 
only recently been pursued in detail\cite{williams95}, is extremely 
difficult due to the direct confrontation with singularities and the 
indefinite norm.  Alternatively, 
Euclidean space is characterized by a positive definite norm, i.e., 
$p^2\equiv p_1^2+p_2^2+p_3^2+p_4^2\ge 0$, and it has long been known that the 
Euclidean formulation is therefore advantageous in the description of 
nonperturbative processes through the Dyson-Schwinger(DSE) and 
Bethe-Salpeter(BSE) equations.  

These components are frequently 
assembled into diagrams such as that in Fig.1 for the calculation of 
form factors\cite{frank94,formfactors}.  The problem with the Euclidean 
formulation of such processes is that for 
physical particles the external momenta must satisfy the mass-shell 
condition $P^2=-M_1^2$ and $K^2=-M_2^2$, thus forcing the return to 
Minkowski space.  One is then faced with the 
problem of complex momenta flowing through the loop in the diagram of Fig.1. 
This in turn requires solving the DSEs in the complex plane.  Although 
some progress has been made in this regard\cite{cahill91}, it is an 
extremely difficult problem and most calculations until now have employed 
entire-function fits on the real axis.  
Although some success has been obtained with this approach\cite{formfactors}, 
there are uncertainties associated with the extrapolation of these functions 
into the complex plane.  
\begin{figure}
\unitlength1.cm
\begin{picture}(14,8)(2.0,-12.0)
\includegraphics{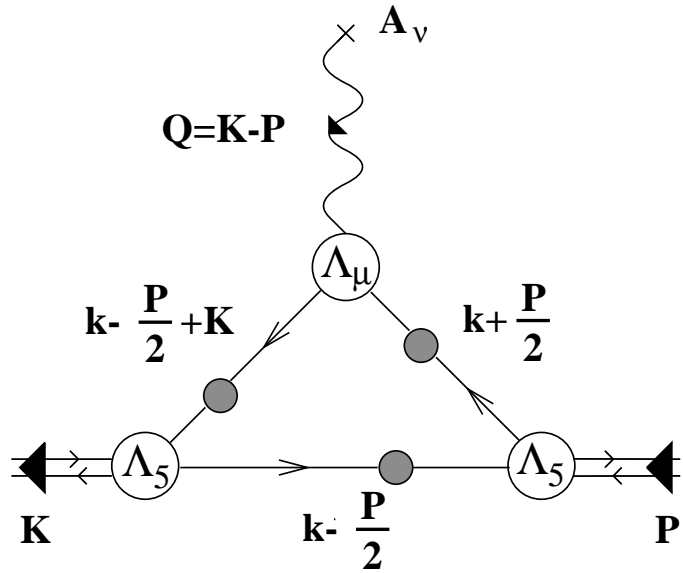}
\end{picture}
\caption{The electromagnetic vertex for a composite pion at tree level.  
The quark Green's 
functions, quark-photon vertex, and the pion Bethe-Salpeter amplitudes, 
are dressed in a 
consistent, gauge-invariance manner.}
\label{fig1}
\end{figure}

Here we offer a fresh approach which avoids these uncertainties by obtaining 
the mass-shell conditions implicitly through an application 
of the Cauchy integral formula.\footnote{A similar method has been applied
in Ref.\cite{roos} to 2-point functions, where it has been used as a means to 
obtain the physical mass. However, in this paper, we would like to demonstrate
that Euclidean time projection methods are much more powerful and can be
successfully applied to higher point functions as well.}  
This approach does not circumvent the 
need to make assumptions about the analytic structure of the components 
in the diagram of Fig.1, but rather relies on the selection of relevant 
singularities in much the same way as is accomplished by 
Euclidean-space lattice calculations.  We consider the pion form factor as 
a prototype for the purposes of illustration.  

The exact calculation of the pion form factor via lattice techniques has 
been studied previously\cite{woloshyn,martinelli}, and proceeds as follows:
one first considers the Euclidean three point correlation function
\begin{equation}
C_\mu(t_x,t_y) = \int d^3x d^3y \langle 0|T\left[
J_\pi^\dagger(y) J_\mu(0) J_\pi(x) \right] |0\rangle e^{-i{\vec K}\cdot {\vec 
y}}
e^{i{\vec P}\cdot {\vec x}},
\label{eq:corr}
\end{equation}
where $J_\pi(x) = \bar{u}(x) i\gamma_5 d(x)$ is the interpolating field
for the pion and $J_\mu(0) =\frac{2}{3}\bar{u}(0) \gamma_\mu u(0)
-\frac{1}{3}\bar{d}(0) \gamma_\mu d(0)$ is
the vector current.  $t_x$ and $t_y$ are the time components of $x$ and $y$
respectively.
Upon inserting a complete set of states between each pair of operators
in Eq.(\ref{eq:corr}) and taking the limit 
$t_x\rightarrow -\infty$ and $t_y \rightarrow 
\infty$,
so that only the lightest state contributes significantly, one finds
\begin{equation}
\lim_{t_y,-t_x\rightarrow \infty}C_\mu(t_x,t_y)=
- \frac{e^{E_{\vec P}t_x} e^{-E_{\vec K}t_y}}{4E_{\vec P}E_{\vec K}}
\langle 0|J_\pi^\dagger(0)| \pi,{\vec K} \rangle
\langle  \pi,{\vec K} |J_\mu(0) |\pi,{\vec P}\rangle
\langle \pi,{\vec P}|J_\pi(0)  |0\rangle ,
\label{eq:tlarge}
\end{equation}
where $|\pi,{\vec P}\rangle $ represents an on-mass-shell (i.e.
$E_{\vec P} \equiv \sqrt{M_\pi^2 + {\vec P}^2}$) pion state with 
three-momentum ${\vec P}$.
Note that Eq.(\ref{eq:tlarge}) is exact. Thus despite the Euclidean formulation
(as long as one does not employ
any further approximations such as quenching, finite quark masses, finite
lattice spacing, finite volume) it allows an exact computation of the
on-shell pion electromagnetic form factor
\begin{equation}
(\tilde{P}_\mu +\tilde{K}_\mu )F_\pi\left((\tilde{P}-\tilde{K})^2\right) = 
\langle  \pi,{\vec K} |J_\mu(0) |\pi,{\vec P}\rangle, 
\end{equation}
where $\tilde{P}\equiv (E_{\vec P},{\vec P})$ and $\tilde{K}\equiv (E_{\vec 
K},{\vec K})$.
In practice \cite{woloshyn} Eq.(\ref{eq:corr}) is usually evaluated
in the quenched approximation, yielding
\begin{eqnarray}
C_\mu(t_x,t_y)= \int dU e^{-S(U)}
\int  && d^3x d^3y e^{-i{\vec K}\cdot {\vec y}}
e^{i{\vec P}\cdot {\vec x}} \nonumber \\
&& \mbox{tr} \left[ \gamma_5 M^{-1}(U)(x,0)
\gamma_\mu M^{-1}(U)(0,y)\gamma_5 M^{-1}(U)(y,x)\right],
\label{eq:quenched}
\end{eqnarray}
where $M^{-1}(U)(y,x)$ is the quark propagator for a given configuration
of link fields $U$. Note that the group integration $\int dU$ implies
that the quark lines and the vertices are dressed with arbitrary gluon
lines.

The above example (plus many other examples in Euclidean lattice gauge theory)
shows that, despite the fact that the calculations are
performed in a Euclidean metric, it is still possible to extract 
on-mass-shell matrix elements, without having to make {\it ad hoc}
assumptions about the analytic continuation back to a Minkowskian
metric. The basic idea is that large Euclidean time merely acts as
a filter to project out the ground state of the Hamiltonian.
In this paper, we present a procedure that uses similar
techniques in the context of Euclidean Dyson-Schwinger calculations,
which also there allows an unambiguous extraction of the on-mass-shell
matrix elements.

\section{The Dyson-Schwinger Approach to Form Factor Calculations}

An alternative to performing the exact (or quenched) calculation 
is to employ two 
approximations restricting the type of allowed gluon dressing.  The first  
is to consider only dressing by the gluon two-point function, and the second 
is to consider only ladder/rainbow dressing.\footnote{Note that ladder/rainbow 
dressing implicitly leads to a restriction on allowed quark lines that resembles
the quenched approximation in lattice QCD.}
These approximations are in fact implied 
by the diagram of Fig.1, and allow the separate calculation of the 
components, i.e., quark propagators and vertex functions.  These 
approximations further comprise an electromagnetic gauge-invariant 
description\cite{frank94,frank95}.  Within this approximation the diagram of 
Fig.1 is given by
\begin{eqnarray}
\Lambda^5_\mu(K,P)=\int \frac{d^4k}{(2\pi)^4}\;\mbox{tr}\Biggl[ && 
\Lambda_5 \left(P,k\right) G\left(k+\frac{P}{2}\right)\Lambda_\mu 
\left(K-P,k+\frac{K}{2}\right) \nonumber \\
&&\times G\left(k-\frac{P}{2}+K\right)
\Lambda_5 \left(K,k-\frac{P}{2}+\frac{K}{2}\right)
G\left(k-\frac{P}{2}\right)\Biggr] ,\label{one}
\end{eqnarray}
where $\Lambda_5$ is the solution of the {\it inhomogeneous} pseudoscalar 
BSE (quark-pseudoscalar vertex), $\Lambda _\mu$ is the solution of the 
{\it inhomogeneous} vector BSE (quark-photon vertex), 
and $G$ is the quark propagator.  

We will use the notation $\Lambda $ 
to represent the solutions to the {\it inhomogeneous} BSE, and 
$\Gamma $ to represent the solutions to the {\it homogeneous} BSE. In the 
vicinity of a pole, they are related by \cite{frank95}
\begin{equation}
\Lambda(P,k)\approx \frac{M^2\sqrt{Z(-M^2)}}{P^2+M^2}\Gamma(P,k),\label{two}
\end{equation}
where $M$ is the pole mass, $Z(-M^2)$ is the wave function renormalization, 
and $\Gamma (P,k)$ therefore obeys the standard 
Bethe-Salpeter bound state normalization\cite{frank95}.  

The vertex function for a photon coupling 
to an on-shell pion is then given by 
\begin{eqnarray}
\Gamma^\pi_\mu(K,P)=\int \frac{d^4k}{(2\pi)^4}\;\mbox{tr}\Biggl[ && 
\Gamma_\pi \left(P,k\right) G\left(k+\frac{P}{2}\right)\Lambda_\mu 
\left(K-P,k+\frac{K}{2}\right) \nonumber \\
&&\times G\left(k-\frac{P}{2}+K\right)
\Gamma_\pi \left(K,k-\frac{P}{2}+\frac{K}{2}\right)
G\left(k-\frac{P}{2}\right)\Biggr] ,\label{three}
\end{eqnarray}
where in the vicinity of the pion pole 
\begin{equation}
\Lambda^5_\mu(K,P)\approx \frac{M_\pi^4Z(-M_\pi^2)}
{(P^2+M_\pi^2)(K^2+M_\pi^2)}\Gamma^\pi_\mu(K,P).\label{four}
\end{equation}
The quantity $\Gamma^\pi_\mu(K,P)$ has the decomposition\cite{frank94} 
\begin{eqnarray}
\Gamma^\pi_\mu(K,P)=\left(P_\mu+K_\mu\right)&&
F_\pi\left(P^2,K^2,(K-P)^2\right)\nonumber \\ 
+&&(K_\mu-P_\mu)(P^2-K^2)H_\pi\left(P^2,K^2,(K-P)^2\right)\label{five}
\end{eqnarray}
and is commonly used to extract the pion form factor, $F_\pi$, at the 
on-shell point $P^2=K^2=-M_\pi^2$.  

Rather than explicitly enforcing the mass-shell condition as has 
been explored previously\cite{formfactors}, here we instead take an 
approach that is 
similar to that taken in the Euclidean-space lattice calculation.  The 
quantity in the DSE approach that is analogous to the three-point 
correlation function, $C_\mu(t_x,t_y)$ in (\ref{eq:quenched}), is 
\begin{equation}
C'_\mu(t_x,t_y)\equiv -\int_{-\infty}^{\infty}\frac{dP_4dK_4}{(2\pi)^2}
\; e^{-iP_4t_x}e^{iK_4t_y}\Lambda^5_\mu(K,P) .\label{dsecor}
\end{equation}
The form factor is then obtained 
implicitly from the expression 
\begin{eqnarray}
\lim_{t_y,-t_x\rightarrow \infty}&&C'_\mu(t_x,t_y)\nonumber \\
=&&-\frac{e^{E_{\vec P}t_x}e^{-E_{\vec K}t_y}}
{4E_{\vec P}E_{\vec K}}\left[M_\pi^2\sqrt{Z(-M_\pi^2)}\right]
\left[(\tilde{P}_\mu+\tilde{K}_\mu)
F_\pi\left( (\tilde{K}-\tilde{P})^2\right)\right]
\left[M_\pi^2\sqrt{Z(-M_\pi^2)}\right].\label{six}
\end{eqnarray}
The correctness of Eq.(\ref{six}) is readily verified by substituting 
Eqs.(\ref{four}) and (\ref{five}) into (\ref{dsecor}), 
along with the implicit assumption 
that there are no poles in the immediate vicinity of the ground-state 
pion pole of interest.  The validity of this assumption is of course 
model dependent.  Eq.(\ref{six}) can be directly compared 
with the exact result in Eq.(\ref{eq:tlarge}).  

\section{Toy-model calculation}

We now proceed to evaluate Eq.(\ref{six}) for a simple model which allows 
direct comparison with the exact result.  That is, we make the following 
choices
\begin{eqnarray}
\Lambda_5(P,k)&\equiv & \frac{i\gamma_5}{P^2}\nonumber \\
\Lambda_\mu(P,k)&\equiv & -i\gamma_\mu \label{seven} \\
G(k)&\equiv &\frac{1}{i\gamma \cdot k+m_q}\nonumber
\end{eqnarray}
where $m_q$ is taken to be a constituent quark mass.  Eq.(\ref{seven}) 
implies in particular that $\Gamma_\pi=i\gamma_5$.  The exact result 
can therefore be obtained by directly fixing the mass-shell condition 
in the evaluation of Eq.(\ref{three}) and Eq.(\ref{five}), and is shown 
by the solid line in Fig.2.  The constituent quark mass, $m_q$, is taken 
to be 300 MeV, and a cutoff of 1 GeV is applied on the relative momentum 
at the quark-photon vertex to regulate the integration.  
The external momenta are constrained such that $P^2=K^2=0$.  

For comparison, we evaluate Eq.(\ref{six}) by substituting the toy model 
(\ref{seven}) into Eq.(\ref{one}) for the vertex $\Lambda ^5_\mu$.  The 
results are shown by the series of dashed curves in Fig.2 for 
time separations $\Delta t\equiv t_y-t_x\sim 2 fm$.  
The Fourier transform is 
performed with 4096 equally-spaced points with a grid spacing of $0.02$ GeV.   
Note that the agreement with the exact result diminishes at larger momentum 
transfers $Q^2$, as is expected.  The agreement at large momentum is improved
by increasing the number of Fourier transform points to include larger 
momentum components in the integration.  

There is an additional effect due to the presence of 
poles in the quark propagators.  This is revealed in Fig.3 where 
for example a constituent quark mass of 1 GeV is used with 1024 
Fourier transform points at a grid spacing of $0.03$ GeV.  With this 
large quark mass the associated singularities are strongly damped and 
therefore do not contaminate the extraction of the pion pole.  
The comparison with the exact result is now quite good out to reasonably large 
momenta.  The increasing discrepancy with increasing time separation at 
large momenta is due to the highly oscillatory nature of the Fourier transform, 
and is improved with a finer grid spacing.  
It is expected that the use of quark propagators from 
DSE studies\cite{frank96} that reflect the confining nature of 
QCD will also produce 
accurate results at higher momenta due to the absence of such poles.  
\begin{figure}
\unitlength1.cm
\begin{picture}(14,20)(2.0,2.0)
\includegraphics{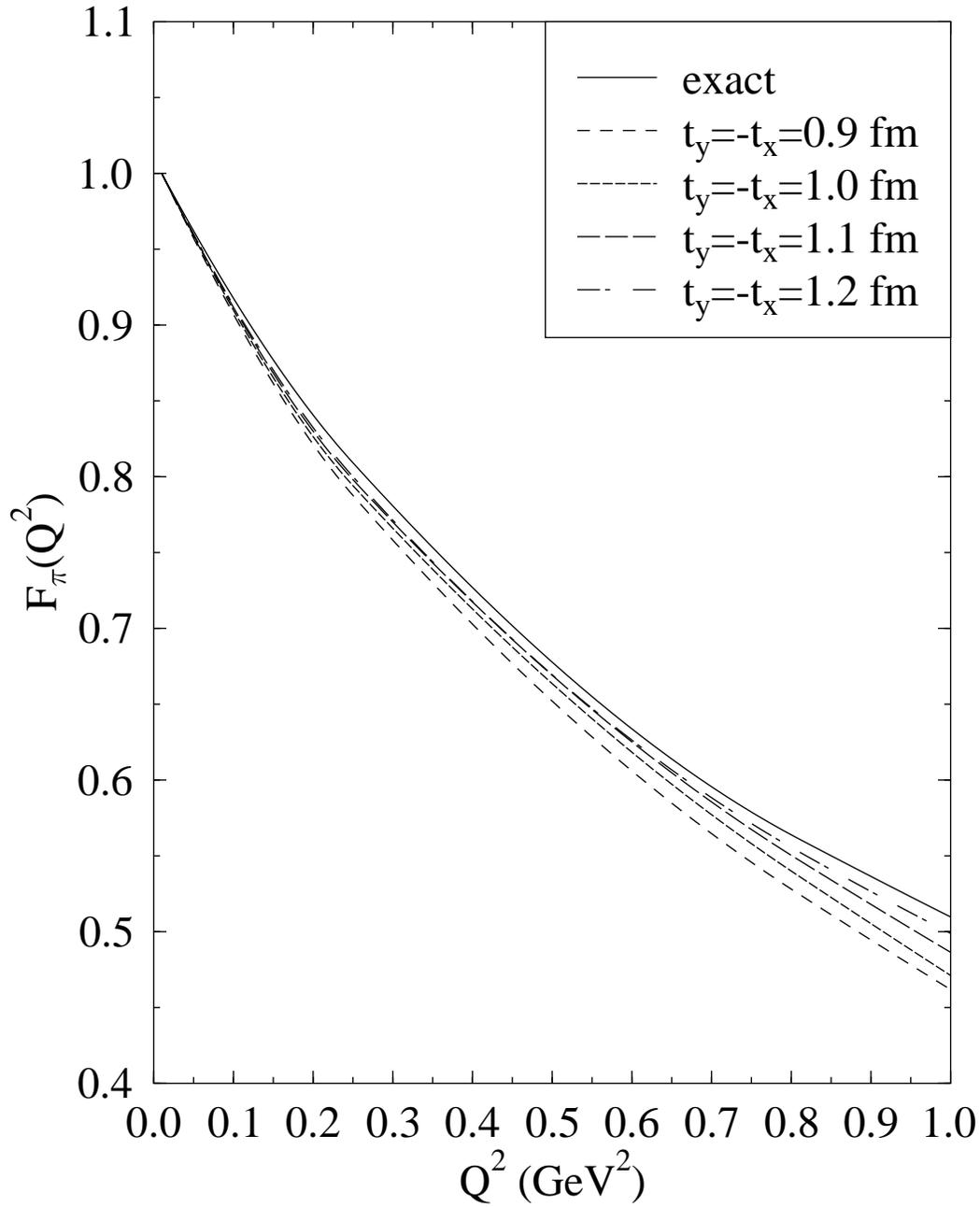}
\end{picture}
\caption{The exact and extracted ``pion" form factors are compared for a 
constituent quark mass $m_q=300$ MeV.}
\label{fig2}
\end{figure}

\begin{figure}
\unitlength1.cm
\begin{picture}(14,20)(2.0,2.0)
\includegraphics{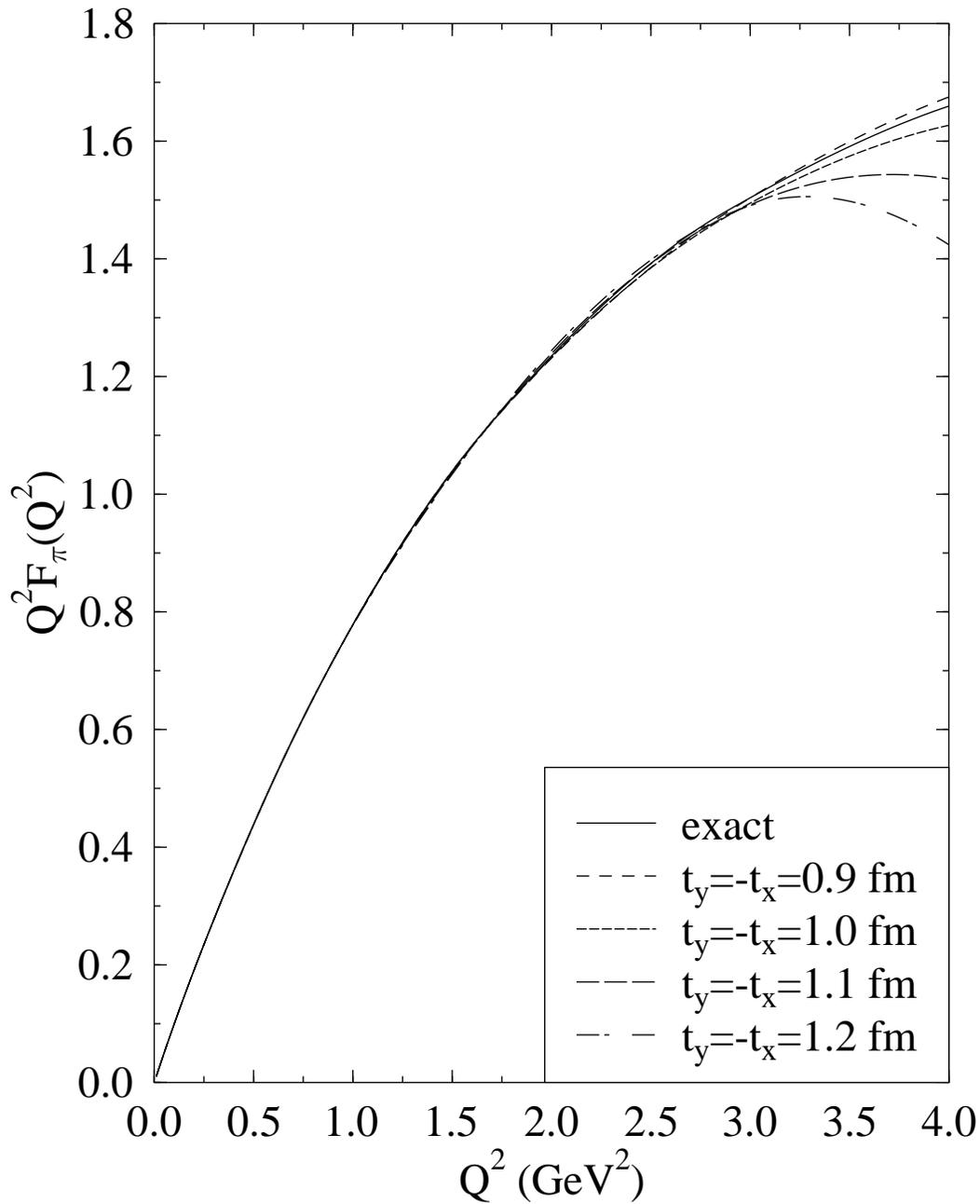}
\end{picture}
\caption{The exact and extracted ``pion" form factors are compared for a 
constituent quark mass $m_q=1.0$ GeV.}
\label{fig3}
\end{figure}

\section{Summary and Outlook}

We have introduced Euclidean time projection methods as a means to project
Euclidean metric 3-point functions onto the physical mass shell.
The method, which is based on Cauchy's theorem and
the existence of a Lehmann representation, avoids having to make {\it ad hoc}
assumptions about the analytic behavior of propagators and vertex functions, 
and thus allows the direct use of Euclidean DSEs 
for the calculation of form factors.  
We have demonstrated the feasibility of the method by studying the
pion form factor for a toy model where the exact form factor is known.
The approach resembles methods that have been used in the context of lattice
gauge theory\cite{woloshyn,martinelli}. The idea is to first Fourier transform 
n-point
functions from (Euclidean) energy to (Euclidean) time. 
The Euclidean time difference between hadron interpolating fields
and currents that are used to ``probe'' the hadron structure is
then taken to be large. This ensures that the hadron which is being
probed is actually the ground state with the quantum numbers
of the interpolating field.

Even though we have studied the pion form factor as a showcase example,
applications of Euclidean time projection methods should be applicable
to a much wider class of physical observables. Starting with
2-point functions, one can obtain masses\cite{roos} as well as
light-cone moments of hadron wave functions\cite{martinelli}
on the mass shell. Examples for 3-point functions that we plan to calculate
include the elastic form factor, moments of parton distribution functions
(e.g. momentum or spin fraction carried by the quarks in a given hadron),
and the $\gamma^* \pi \rightarrow \gamma$ transition form factor. The most 
simple
observables related to 4-point functions that one might consider calculating
using Euclidean time projection are hadron polarizabilities and (virtual)
Compton scattering processes.

The main limitations of the method are first of all the restriction to
the lightest hadrons for given quantum numbers: though theoretically
possible, it is in practice rather difficult to
use the method to obtain observables that involve excited states, since
the whole trick is to use large Euclidean times to suppress excited states
(a limitation that is very familiar to the lattice gauge community).
Another limitation concerns very large momentum transfers, where rapid
oscillations of the integrand require a large number of integrations
points. However, even with rather moderate computing power, this still
allows access to most of the kinematic range that is available at TJNAF.

\acknowledgements

MB would like to thank the Department of Energy (contract DE-FG03-96ER40965) 
as well as TJNAF for support.  MRF is supported by the Department of 
Energy under Grant \# DE-FG06-90ER40561.  KLM is supported by 
the Natural Sciences and Engineering Research Council of Canada.

\end{document}